## Citation

**MLA Style:** Maruthi Rohit Ayyagari *"Cache Contention on Multicore Systems An Ontology-based Approach"* International Journal of Engineering Trends and Technology 67.5 (2019): 58-62.

**APA Style:** Maruthi Rohit Ayyagari (2019). Cache Contention on Multicore Systems An Ontology-based Approach *International Journal of Engineering Trends and Technology*, 67(5), 58-62.

```
@article{Ayyagari2019,
author = {Ayyagari, Maruthi Rohit},
journal = {International Journal of Computer Trends and Technology (IJCTT)},
number = {5},
pages = {58--62},
title = {{Cache Contention on Multicore Systems: An Ontology-based Approach}},
volume = {67},
year = {2019}
}
```

# Cache Contention on Multicore Systems An Ontology-based Approach

Maruthi Rohit Ayyagari

*College of Business, University of Dallas, Texas, USA*
E-mail: rayyagari@udallas.edu

*Abstract*— *Multicore processors have proved to be the right choice for both desktop and server systems because it can support high performance with an acceptable budget expenditure. In this work, we have compared several works in cache contention and found that such works have identified several techniques for cache contention other than cache size including FSB, Memory Controller and prefetching hardware. We found that Distributed Intensity Online (DIO) is a very promising cache contention algorithm since it can achieve up to 2% from the optimal technique. Moreover, we propose a new framework for cache contention based on resource ontologies. In which ontologies instances will be used for communication between diverse processes instead of grasping schedules based on hardware.*

*Keywords- component; multicore; cache; contention; FSB; ontology*

## I. Introduction

According to Moore law, system speed is going to increases dramatically in 18-24 months [1]. As a result, multicore processors have become so dominant for both desktops and servers because it can give more performance compared to traditional systems. Advanced multicore computing systems usually share caches to support data sharing and allow fast communication. The most important cache is the last level cache (LLC) which is being shared by more than one core, usually two cores.

Although the LLC will allow fast communication between cores, the cache can be contended by different cores. In this case, the system will need to read data and instructions from the main memory and fetch it back to the cache which is considered a time-consuming process compared to the speed of the cores. This process is referred to as a cache miss and is very painful to the application that requires high Quality of Service (QoS) such as could computing environment.

To understand how to cache contention can occur, Figure 1 illustrates a scenario. Assume we are using c0 and c1 cores, and if two threads of the same application are run together, then it is possible to have cache contention on the L2 cache (referred to as intra-application contention). If different applications threads run then a possible cache contention on L2, is also conceivable. Another type of contention called Front Side Bus (FSB) contention might occur if different threads are run on different cores that do not share the same L2 cache, for example, c0 and c3 cores.

There have been many approaches to resolve or reduce such cache contention, most of these approaches depend on three major components which are: the used benchmark, the classification scheme, the policies, and the used algorithms. The NAS Parallel Benchmarks (NPB) [2], [3], The Standard Performance Evaluation Corporation (SPEC) [4], [5] and Princeton Application Repository for Shared-Memory Computers (PARSEC) [6] are among the popular benchmarking software. Schemes are used for finding the best co-runner of an application with another. Scheduling algorithms implement the policies to assign threads to cores given the application classification. There are two categories of such algorithms, one is online, and the other is offline. The online algorithm gets cache statistics using the performance counter [7], while the offline algorithms use a prediction approach to calculate statistics offline before performing the benchmark, such as the total cache access in a specific number of millions of transactions [8].

To the best of our knowledge, none of the proposed solutions to the cache contention gives a holistic view of all the possible factors that might affect cache contention performance. This work tries to give an insight into this direction. The contribution of this work is 1) compare various classification schemes and used policies 2) compare famously used algorithms namely: Utility Cache Partition [9], Stack Distance Profile [10], Page Coloring [11], Static and dynamic Static Scheduling Order Adjustment (SOA) [12], Distributed Intensity (DI) [13], and Distributed Intensity Online (DIO) [13]; and 3) propose a model to cache contention based on an ontology.

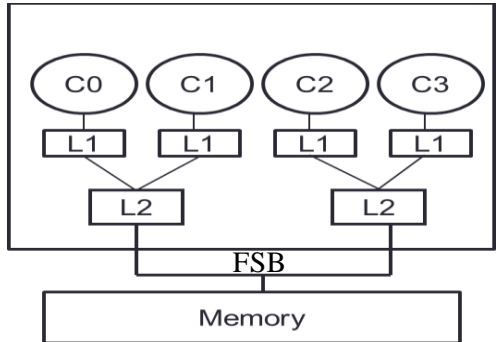

Fig 1. Illustration of cache contention.

The paper is organized as follows. Section II describes Schemes and policies. The comparison between algorithms is conducted in Section III. Our proposed model is in Section IV. We discuss the results in Section V. Related work is in Section VI. VII concludes the paper.

## II. CLASSIFICATION SCHEMES AND POLICIES

### A. Classification schemes

Schemes are used for finding which threads should and should not be scheduled together. It is very crucial to select application threads such that they both execute together and the cache minimally contends. On the one hand, if two application is cash busters, we will end up with deficient performance due to high cache misses. On the other hand, if the two applications are using the low cache, then the cache is wasted.

**Miss Rate:** A cache miss denotes a failed try to read or write cached data, which in turn causes a high delay. Cache misses can be due to instruction, data read, or write misses. A read miss from an instruction cache frequently causes the highest delay, because the executed thread, has to stall until the instruction is fetched from memory. Research has been conducted on cache behavior in an attempt to find the best combination of cache size with other factors. One substantial contribution by Mark Hill [14], is the separation of misses into three categories: Compulsory misses initiated by the first reference to a datum, Capacity misses are those misses that occur due to the limited size of the cache associativity, and Conflict misses which are misses that could have been escaped, if the cache did not remove an entry before. Another work by Knauerhase et al. found the best association between two applications threads [15].

**Animal Class Scheme:** Dynamic cache partitioning scheme that performs marginally enhanced than other schemes such as Utility Cache Page Partition (UCP) while incurring a lower implementation cost. It depends on heuristics based on animal classification [16]: Turtles: low cache usage; Sheep: miss rate is low, but it is insensitive to the number of cache ways allotted to it; Rabbits: miss rate is low but is sensitive to several apportioned cache ways — the Tasmanian Devils[16] — threads with high miss rate, and low cache reuse.

**Pain Scheme:** A Scheme based on two concepts of cache sensitivity and cache intensity. The cache sensitivity is the amount of how much an application will feel pain when cache space is taken from it due to the contention while intensity refers to the quantity of how much a thread will hurt others by taking space from them, in a shared cache [13]. These values are formally calculated and summed relative to each other, to calculate the effect of both to each other.

**"Perfect" Scheduling Scheme (Optimal Scheme)**: This scheme constructs a graph representation of the threads or applications that need to execute (co-run together) at some point in time. Threads are signified as nodes linked by edges; edges weights are evaluated by the sum of the joint co-run degradations between two threads. The optimal scheduling job can be found by solving the graph minimization problem [17]

### B. policies

A scheduling policy is the set of decisions made regarding cache scheduling priorities. A scheduling algorithm is the instructions that implement a given scheduling policy. There are different policies aside from DEFAULT Linux policy such as:

**"Perfect" Scheduling Policy**: Jiang's way for outlining the optimal and the worst thread schedule policy[17].

**Greedy Policy**: a process is selected in the slave CPU to couple with the current process on the master CPU such that the effect to the process on the master CPU is the minimum.

**Statistical Policy**: a policy which depends on recorded history data to make the co-runner selection as accurate as possible. This type of policy is more accurate than a greedy policy, but it has more overhead since more data structures have to be implemented to support information history storage, such as an array of PIDs [18].

**Stall Cycles Policy**: In this policy, the stall(wait) cycles are used to select co-runners. Co-runners are selected such that they have the most significant difference in stall cycles. Under this selection, the tasks with different performance will undoubtedly be co-scheduled together.[19]

**Centralized Sort**: Application threads lists are sorted by the miss rate value, and then they are allocated to cores in order. In this policy, the total miss rate of allocated is flattened across every cache[20].

## III. EVALUATION OF CACHE CONTENTION ALGORITHMS

A scheduling algorithm is the programming code that implements a given scheduling policy. There have been many algorithms in operating system community in order to resolve cache contention. We categorize them as offline or online algorithms. The online algorithm gets cache statistics using the performance counter, while offline ones uses a prediction approaches (the number of last level cache access per one million transactions) to calculate statistics offline before performing the benchmark. Table 1 is a comparison of some of these algorithms.

**Cache Page Coloring Algorithm**: This algorithm enforces page coloring on only a small set of frequently accessed pages for each process. The cost of identifying hot pages online is reduced by leveraging the knowledge of spatial locality during a page table scan of access bits. [11]

**Utility Cache Page Partition (UCP):** a custom hardware solution that estimates each thread's number of hits and misses for all possible number of ways assigned to the application in the cache built on stack-distance profiles. The cache is then partitioned for the co-running applications to cut down the number of cache misses. This algorithm decreases cache contention once co-runners are known ahead of time. [9]

Table1- Algorithms comparison

| Measure / Algorithm | Contention Measurement | Policy | Scheme | Dependency |
|---|---|---|---|---|
| **Cache Page Coloring** [11] | LLC | Cache page coloring | N/A | - |
| **Cache Partition** [9] | LLC | - | - | Hardware |
| **Stack Distance Profile** [10] | LLC | - | - | Stack profile |
| **DI** [13] | L1, L2, memory controller | Centralized order | Miss rate, pain | Stack profile |
| **DIO** [13] | L1, L2, memory controller | Centralized order | Miss rate, pain | Stack profile |
| **Static SOA** [12] | L1, L2, FSB | Stall cycles | - | - |
| **Dynamic SOA** [12] | L1, L2, FSB | Stall cycles | - | - |

**Distributed Intensity (DI) and DI Online (DIO):** DI assign threads to the solo miss rate as found by the stack distance profile algorithm. Then applications are distributed through caches such that the miss rates are distributed as consistently as probable. DIO is built on the same classification scheme and scheduling policies as DI, but it obtains the miss rates of threads online via performance counters.

**Static and Dynamic SOA**: The idea of these algorithms is to reduce cache contention by adjusting the scheduling order of threads to execute appropriately. The Static Scheduling Order Adjustment (SOA) method acquires the cache requirement information statically by offline profiling.

## IV. PROPOSED FRAMEWORK

We propose to use an ontology for resolving shared cache contention. Ontologies have proven to be one way of ensuring various mapping systems [21][22]. Moreover, ontologies were invented as one way to resolve the problem of interoperability so we can utilize this idea in resolving cache contention problems [23]. The nature of the cache contention problem reveals to be possibly solved by an ontology. Since the current solutions to cache contention are application and architecture specific, we believe an ontology-based approach might give some insight into a possible solution.

Figure 2, shows our proposed model. We want each thread to have the ability to update the instance of the ontology so that we can integrate different processes or threads regardless of the system architecture. The ontology

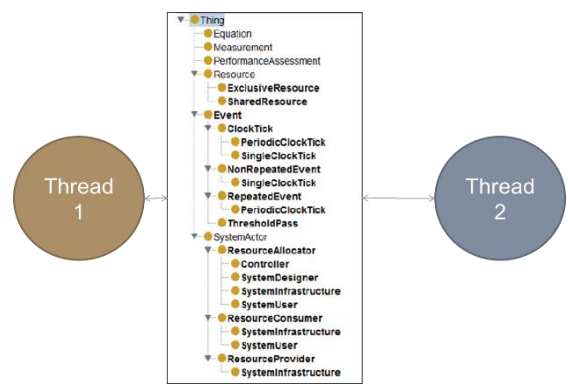

Fig 2. cache contention ontology model

class in Figure 2 is from the work of Rohloff [24]. However, further refinement to this class model is needed in order to have a model that match cache contention requirements [25], using datamining techniques [26], and integrating software engineering techniques [27][28].

## V. DISCUSSION

Surveyed works have identified several reasons for cache contention including cache space, FSB, Memory Controller, prefetching hardware. DIO perform within 2% of the optimal [29]. The highest impact of contention-aware scheduling procedures is in improving the quality of service or performance isolation for individual applications. Front Side Bus is a significant factor of the benchmarks and degrades performance by more than 11% [12]. For the dynamic SOA method, the execution time reduction can achieve up to 7.09% [18].

We noticed that the surveyed algorithms are application and architecture dependent. Moreover, each one has its own benchmark which is relatively very difficult to conduct a real comparison between them. Therefore, a holistic review is needed to find an efficient algorithm.

Therefore, we propose a new framework for cache contention based on resource ontologies. In which ontologies instances will be used for communication between diverse processes instead of grasping schedules based on hardware.

## VI. Related work

Although there has been tremendous growth in the use of ontologies to facilitate systems and service integration in general , there has been little work on general ontologies for the critical challenge of resource sharing as needed for offline or online resource allocation and reallocation. [23], [30]–[32].

Probably the nearest work to our work is the one introduced by Rohloff [24]. However, it was not intended for cache contention, and it was introduced for any resource sharing for the goal of integration.

## VII. Conclusion

Several works have identified several reasons for cache contention other than cache size including: FSB, Memory Controller, prefetching hardware. DIO perform within 2% of the optimal. The highest impact of contention-aware scheduling techniques is in improving the quality of service or performance isolation for individual applications. Front Side Bus is a significant factor of the benchmarks and degrades performance by more than 11%. For the dynamic SOA method, the execution time reduction can achieve up to 7.09%. We propose a new model for cache contention based on ontologies. In which ontologies instances will be used for communication between diverse processes instead of mastering schedules based on hardware. Our model still needs further research to quantify its parameters. The proposed approach can resolve the problems of architecture dependent technique by using a shared resource. In the future, the proposed technique will be elaborated and examined.

## References

[1] V. Transcript, "Excerpts from a Conversation with Gordon Moore: Moore" s Law," *Intel Corp.*, 2005.

[2] D. Bailey, T. Harris, W. Saphir, R. Van Der Wijngaart, A. Woo, and M. Yarrow, "The NAS parallel benchmarks 2.0," 1995.

[3] D. H. Bailey, "The NAS Parallel Benchmarks: History and Impact," 2015.

[4] M. Dey, A. Nazari, A. Zajic, and M. Prvulovic, "EMPROF: Memory Profiling Via EM-Emanation in IoT and Hand-Held Devices," in *2018 51st Annual IEEE/ACM International Symposium on Microarchitecture (MICRO)*, 2018, pp. 881–893.

[5] J. L. Henning, "SPEC CPU2000: Measuring CPU performance in the new millennium," *Computer (Long. Beach. Calif).*, vol. 33, no. 7, pp. 28–35, 2000.

[6] X. Fu et al., "New parsec data base of $\alpha$-enhanced stellar evolutionary tracks and isochrones--I. Calibration with 47 Tuc (NGC 104) and the improvement on RGB bump," *Mon. Not. R. Astron. Soc.*, vol. 476, no. 1, pp. 496–511, 2018.

[7] R. West, P. Zaroo, C. A. Waldspurger, and X. Zhang, "Online cache modeling for commodity multicore processors," *ACM SIGOPS Oper. Syst. Rev.*, vol. 44, no. 4, pp. 19–29, 2010.

[8] G. Liu, J. Park, and D. Marculescu, "Dynamic thread mapping for high-performance, power-efficient heterogeneous many-core systems," in *2013 IEEE 31st international conference on computer design (ICCD)*, 2013, pp. 54–61.

[9] M. Qureshi, "Utility-based cache partitioning: A low-overhead, high-performance, runtime mechanism to partition shared caches," *Proc. 39th Annu. IEEE/ACM*, 2006.

[10] D. Chandra, F. Guo, and S. Kim, "Predicting inter-thread cache contention on a chip multi-processor architecture," *Comput. Archit.*, pp. 340–351, 2005.

[11] X. Zhang, S. Dwarkadas, and K. Shen, "Towards practical page coloring-based multicore cache management," in *Proceedings of the 4th ACM European conference on Computer systems*, 2009, pp. 89–102.

[12] T. Dey, W. Wang, and J. W. Davidson, "Characterizing multi-threaded applications based on shared-resource contention," *Syst. Softw.*, pp. 76–86, 2011.

[13] S. Zhuravlev and S. Blagodurov, "Addressing shared resource contention in multicore processors via scheduling," *ACM SIGARCH Comput.*, vol. 38, no. 1, pp. 129–142, 2010.

[14] M. D. Hill and A. J. Smith, "Evaluating associativity in CPU caches," *Comput. IEEE Trans.*, vol. 38, no. 12, pp. 1612–1630, 1989.

[15] R. Knauerhase, P. Brett, B. Hohlt, and T. Li, "Using OS observations to improve performance in multicore systems," *Micro, IEEE*, pp. 54–66, 2008.

[16] Y. Xie and G. H. Loh, "Dynamic Classification of Program Memory Behaviors in CMPs," *Cmp-Msi'08*, no. June, pp. 1–9, 2008.

[17] Y. Jiang, X. Shen, and J. Chen, "Analysis and approximation of optimal co-scheduling on chip multiprocessors," *Proc. 17th*, 2008.

[18] Y. Wang, Y. Cui, P. Tao, H. Fan, Y. Chen, and Y. Shi, "Reducing Shared Cache Contention by Scheduling Order Adjustment on Commodity Multi-cores," *2011 IEEE Int. Symp. Parallel Distrib. Process. Work. Phd Forum*, pp. 984–992, May 2011.

[19] S. Kumar and P. K. Singh, "An overview of modern cache memory and performance analysis of replacement policies," in *2016 IEEE International Conference on Engineering and Technology (ICETECH)*, 2016, pp. 210–214.

[20] M. Tawarmalani, K. Kannan, and P. De, "Allocating objects in a network of caches: Centralized and decentralized analyses," *Manage. Sci.*, vol. 55, no. 1, pp. 132–147, 2009.

[21] I. Atoum, A. Otoom, and A. Abu Ali, "Holistic Cyber Security Implementation Frameworks: A Case Study of Jordan," *International Journal of Information, Business and Management*, vol. 9, no. 1. Elite Hall Publishing House, Taiwan, Republic of China, pp. 108–118, 2017.

[22] I. Atoum and A. Otoom, "Mining Software Quality from Software Reviews: Research Trends and Open Issues," *International Journal of Computer Trends and Technology (IJCTT)*, vol. 31, no. 2. Seventh Sense Research GroupTM, pp. 74–83, 2016.

[23] D. Smirnov and P. Stutz, "Use case driven approach for ontology-based modeling of reconnaissance resources on-board UAVs using OWL," in *2017 IEEE Aerospace Conference*, 2017, pp. 1–17.

[24] K. Rohloff and J. Loyall, "An Ontology for Resource Sharing," in *Semantic Computing (ICSC), 2011 Fifth IEEE International Conference on*, 2011, pp. 530–537.

[25] G. Vandita and G. Sugandha, "Service Differentiation based on Contention Window with Enhanced Collision Resolution LR-WPANs," *Int. J. Comput. Trends Technol.*, vol. 19, no. 2, pp. 86–90, 2015.

[26] M. R. Ayyagari, "Integrating Association Rules with Decision Trees in ObjectRelational Databases," *Int. J. Comput. Trends Technol.*, vol. 67, no. 3, pp. 102–108, 2019.

[27] M. R. Ayyagari and I. Atoum, "CMMI-DEV Implementation Simplified : A Spiral Software Model," *Int. J. Adv. Comput. Sci. Appl.*, vol. 10, no. 4, pp. 445–450, 2019.

[28] M. R. Ayyagari, "iScrum: Effective Innovation Steering using Scrum Methodology," *Int. J. Comput. Appl.*, vol. 178, no. 10, pp. 8–13, May 2019.

[29] S. Zhuravlev and J. C. Saez, *Survey of Scheduling Techniques for Addressing Shared Resources in Multicore Processors*, vol. V, no. September. 2011.

[30] M. Mao, Y. Peng, and M. Spring, "Ontology mapping: as a binary classification problem," *Concurr. Comput. Pract. Exp.*, vol. 23, no. 9, pp. 1010–1025, Dec. 2011.

[31] J. Davies, R. Studer, and P. Warren, *Semantic Web Technologies: Trends and Research in Ontology-based Systems*, vol. 3, no. 1. John Wiley & Sons, 2006.

[32] J. Lehmann, S. Auer, S. Tramp, and others, "Class expression learning for ontology engineering," *Web Semant. Sci. Serv. Agents World Wide Web*, 2011.